\newlength{\back}
\newlength{\slashwidth}
\def\g5{\gamma_5}

\def\Tr{\mbox{Tr}}
\def\tr{\mbox{tr}}
\def\dag#1{#1^{\dagger}}

\def\bea{\begin{eqnarray}}
\def\eea{\end{eqnarray}}
\def\bra#1{\big<#1\big|}
\def\ket#1{\big|#1\big>}
\def\bracket#1#2{\big<#1\big|#2\big>}

\def\bd{\begin{description}}
\def\ed{\end{description}}
\def\ben{\begin{enumerate}}
\def\een{\end{enumerate}}

\documentstyle[12pt]{article}

\begin{document}
\newcommand{\fslash}[1]{{\mbox{$\!\not\!#1$}}}
\newcommand{\refer}[1]{(\ref{#1})}
\newcommand{\bold}[1]{\mbox{\boldmath ${#1}$}}
\baselineskip 4 ex

\title{Elimination of the Landau Ghost from Chiral Solitons}

\author{Josef Hartmann$^1$
 and Friedrich Beck \\
        Institut fuer Kernphysik \\
        Technical University of Darmstadt \\
        Schlo\ss gartenstr. 9, D-64289 Darmstadt, Germany\\
        {  }\\
        and\\
        {  }\\
        Wolfgang Bentz \\
        Department of Physics, \\
        Faculty of Science \\
        University of Tokyo \\
        Hongo 7-3-1, Bunkyo-ku, Tokyo 113, Japan}

\footnotetext[1]{Supported by
 {\it Deutsche Forschungsgemeinschaft} under contract BE 348/11--1,2}

\date{}
\maketitle
\begin{abstract}
We show a practical way based on the K\"{a}ll\'{e}n-Lehmann representation
for the two-point functions to eliminate the instability of the vacuum against
formation of small sized meson configurations in the chiral $\sigma$ model.

\end{abstract}

\newpage
\section{Introduction}
\par
Since QCD is presently intractable at low energies, approximation schemes
based on effective models have been introduced \cite{EFF}. These models in a
way
incorporate some important features of QCD, like chiral symmetry and its
spontaneous breakdown.
\par
Though being approximative, these effective models still cannot be treated
exactly. Therefore we have essentially two levels of approximation: i)
when we choose the model itself and ii) when we specify an approximation
scheme in order to perform actual calculations within the model.
This implies that the results may depend on the actual approximation scheme.
Therefore, a failure of the model could be due to the model itself, due to
the approximation scheme or both. So, if we want to test the model, we should
be careful about the approximation scheme and keep the dependence on ad hoc
parameters as small as possible.
\par
One of the effective models of QCD to describe the nucleon is the chiral
$\sigma$-model \cite{GL}. The standard procedure to calculate
nucleonic properties in the soliton picture
is to use the loop expansion up to the one quark loop level, i.e; to
integrate out the quark degrees of freedom and use the remaining purely
mesonic effective action at the zero loop level in the mesonic fields
\cite{ACT}.
One can choose some cut-off procedure to regularize the quark determinant
\cite{BR},
but in any case one has to renormalize by
introducing appropriate counter
terms in order to define the parameters appearing in the original Lagrangian
of the model \cite{REN}. Eventually one can let the cut-off
go to infinity as it is
usually done in renormalization theory, or leave it finite.
We note that in the $\sigma$-model, however, neither the cut-off scheme
itself nor the size of the cut-off can be determined uniquely.
\par
It is known, however, that the loop expansion for asymptotically non free
theories without ad hoc cut-off parameters displays an unphysical pole in
the corresponding propagators, the so called Landau ghost \cite{LAN}.
Therefore, if
one follows the procedure outlined above, one finds that the renormalized
$\sigma$-model is plagued by the Landau ghost,
which leads to an instability of the usual translational invariant vacuum
\cite{RK,VAC}.
In view of our remarks at the beginning of this section,
the problem of this Landau ghost can be treated in several ways:
\begin{enumerate}
\item The effective model is considered as unphysical and is abandoned.
\item The approximation scheme (loop expansion) is considered appropriate,
but the effective model can be used only with a regularization
scheme, hereby being restricted to energies far below the unphysical pole.
\item The approximation scheme is improved such that the effective model
does not a priori yield unphysical results for energies up to the
order of the baryon mass.
\end{enumerate}
The second viewpoint implies that vacuum loops are either simply
discarded, or calculated with finite cut-off.
Concerning the first alternative (drop vacuum loops), we note that in some
situations, like in an external field, such a treatment violates conservation
laws \cite{RPA} and cannot be implemented consistently. Besides this,
the investigation of the change of the vacuum structure due to
the finite density should be an important subject for any relativistic
field theory. Concerning the
second alternative (finite cut-off), we note that the Landau ghost
in hardonic theories occurs at rather low energies of the order of the
baryon mass ($\simeq 1 GeV$). This means that its presence is
a real problem for effective quark theories, and its avoidance by the
introduction of an ad hoc parameter is somewhat unsatisfactory.
Since the energy scale
relevant for effective quark theories is set by hadronic masses, a treatment
according to the third viewpoint, which we will adopt in this paper,
is a feasible alternative. We consider the
Landau ghost as one of the basic infinities in effective field theories
which should be brought under control without the introduction of further
parameters. In a different context, this viewpoint has been taken already
a long time ago by Redmond and Bogoliubov et al \cite{RED}
who have shown how to construct
ghost free propagators in the framework of the loop expansion
based on the requirement of the
K\"{a}ll\'{e}n-Lehmann representation.
Their propagators have several interesting features: The associated
wave function renormalization constant is finite, and they have an essential
singularity at $g^2=0$ ($g$ is the coupling constant) which is expected
from more intuitive physical arguments \cite{DYS}. Moreover, since the
associated running coupling constant has no singularity at space-like
momentum transfers, the 'zero charge problem' is avoided.
Their method has recently been
formulated in the language of the effective action \cite{TAN},
which is suitable for
effective quark theories, and also applied to infinite systems \cite{TAN,KA2}.
The main
purpose of the present paper is to show a practical way to incorporate
the ghost elimination in a finite system like the soliton, and to
demonstrate that with this procedure the translational invariant vacuum
becomes stable with respect to decay into an array of small sized
configurations. We will {\em not} construct the fully
self consistent solutions including valence quarks in this paper
which is left for a future work.
(For another possibility to avoid the Landau ghost by introducing vector
bosons, see ref. \cite{GAUGE}.)
\par
The $\sigma$-model has so far not yet been fully exploited to obtain
self consistent solutions including the effect of the Dirac sea. One of
our motivations for investigating this model is as follows: In contrast to
the Nambu-Jona-Lasinio (NJL) model \cite{NJL}, which has been
extensively used to construct solitonic solutions \cite{SOLI,KA1,ISH},
it has a $\phi^4$ interaction term with variable
strength (specified by the $\sigma$ mass) which controls the deviation of the
chiral radius from its vacuum value. Such a term would contribute to the
ground state energy the more the farther the chiral fields are away
from the chiral circle of the vacuum. Since it is known that interactions
terms of the form $\phi^4$ \cite{PHI} or of higher order like
the t'Hooft determinant interaction \cite{KA1} can prevent the
collapse of the NJL soliton, there is the
possibility that stable solitons exist for the linear $\sigma$-model.
The major obstacle to a full investigation of the solitonic sector of the
$\sigma$-model, however, was the instability of the translational
invariant vacuum. The fact, to be established in this paper, that the
method of Redmond and Bogoliubov et al leads to a stabilization of the
translational invariant vacuum is a first step towards self consistent
solutions in the $\sigma$-model.
\par
The rest of the paper is organized as follows: To be self contained and to
set the formalism we state in sect. 2 some results for the ground
state energy in the $\sigma$-model obtained earlier \cite{RK}, and explain
the role of the Landau ghost in a way which seems most transparent to us
and which is appropriate for implementing the ghost elimination procedure.
In sect. 3 we construct
a ghost free model and show results for the ground state energy, and in
sect. 4 we summarize our results.

\section{The ground state energy and the
Landau ghost}

We now proceed to exhibit how the Landau ghost contributes to the
vacuum instability.
The Euclidean Lagrangian of the $\sigma$-model reads, after introducing
the 'physical' parameters $m_{\sigma}$ and $m_{\pi}$ instead of the original
ones $\mu^2$ and $\lambda^2$ \cite{GL}:
\begin{equation}
{\cal L}   =  {\cal L}_F+{\cal L}_M+{\cal L}_{SB}
\label{eq:lag}
\end{equation}
with
\begin{eqnarray}
&{\cal L}_F   =  \overline {\psi} D \psi; \hspace{1.5 cm}
                   D=-i\fslash{\partial}+MU \\
\label{eq:l1}
&{\cal L}_M   =  \frac{v^2}{2}\left[\left(\partial_{\mu}U\right)
                               \left(\partial_{\mu} U^+ \right)
    +\frac{1}{4}(m_{\sigma}^2-m_{\pi}^2) \left(U^+U-1\right)^2
    +m_{\pi}^2 \left(U^+U-1\right)\right].
\label{eq:lmes}
\end{eqnarray}
Here $\psi$ stands for the u and d quarks, $M=gv$ is the
effective quark mass,
${\displaystyle U=\frac{1}{v}(\sigma+i\gamma_5{\bold \pi}\cdot{\bold \tau})}$
is the chiral field, and ${\cal L}_{SB}=f_{\pi}m_{\pi}^2\,\sigma$ with
$f_{\pi}$ the pion decay constant. The model
contains the two free parameters $g$ (or $M$) and $m_{\sigma}$. $v$ is the
vacuum expectation value of $\sigma$, and in the symmetric limit
($m_{\pi}=0$) we have $v=f_{\pi}$\cite{GL}.
Our Euclidean metric is such that $x_{\mu}=x^{\mu}=(\tau,{\bf r})$
with $\tau=ix^0$, $0<\tau<\beta$ with $\beta$ the upper limit of
Euclidean time integration, and $\gamma_{\mu}=\gamma^{\mu}
=(i\beta_D,{\bold \gamma})$ with $\beta_D$ the usual Dirac
$\beta$-matrix. Defining the effective
bosonic action $\Gamma$ from the generating functional $Z$ as usual
\cite{ACT} by
$Z=\int DU \exp(-\Gamma)$ we obtain after integrating out the quark fields
\begin{equation}
\Gamma =  \Gamma_F + \Gamma_M +\Gamma_{SB}
\label{eq:sum}
\end{equation}
with \cite{RK}
\begin{eqnarray}
&\Gamma_F =  -\frac{1}{2} Tr\, \ln D^+D + c.t.
      =  -\frac{1}{2}Tr\, \ln  (1+GV)+\frac{1}{2}Tr\, GV
     -\frac{1}{4}Tr\, G^2V^2  \label{eq:gferm}\\
&\Gamma_M =  \int d^4 x \,{\cal L}_M\,;\,\,\,\,\,\,\,\,
\Gamma_{SB} =  \int d^4 x \,{\cal L}_{SB}.
\label{eq:gmes}
\end{eqnarray}
Here $\Tr A \equiv N_C\int d^4 x\, \tr \bra{x}A\ket{x}$ where $N_C=3$
is the number of colors and $tr$ refers to the Dirac and isospin indices,
$G=(-\partial^2+M^2)^{-1}$, and
\begin{equation}
V = iM \fslash{\partial} U + M^2(U^+U-1).
\label{eq:V}
\end{equation}
The integrals over the Euclidean time $x_0$ extend from 0 to $\beta$.
The counter terms (c.t.) in \refer{eq:gferm}, which include the
subtraction
of the translational invariant vacuum contribution ($U=1$), are determined
such that $\Gamma_F$ gives no terms of the same form as those already
present in $\Gamma_M$.
This corresponds to the mesonic mass and wave function renormalization
at the renormalization point $\mu^2=0$ \cite{REN}. From here on we
consider the chiral
symmetric case ($m_{\pi}=0$).
\par
The vacuum instability can be seen by expanding the ground state energy
$E =\Gamma/\beta = (\Gamma_F + \Gamma_M)/\beta$ in powers of a
characteristic length scale $R$ where the classical
meson fields are localized. We assume time independent fields
$U({\bf r}) = \tilde {U}({\bf x})$ with
${\bf x}={\bf r}/R$. If we use ${\bf x}$ as the integration variable,
every derivative gives rise to a factor
$1/R$, and therefore terms involving derivatives will be
dominant for small R compared
to those without derivatives. Therefore, up to $O(R)$  $\Gamma_M$
contributes only the kinetic term.
To obtain the lowest order contribution from $\Gamma_F$ we
expand the logarithm in \refer{eq:gferm} and note that
the term linear in $V$ is cancelled by a counter term.
Then for small R the leading term is the one quadratic in the
derivatives, i.e; quadratic in the first term of eq. \refer{eq:V}. From the
above we see that the linear and the
non linear model look the same at small sizes. The complete term
of order $V^2$ is
\begin{equation}
\Gamma_F^{(s.s)}=\frac{1}{4} \Tr  (GVGV-G^2V^2) .
\label{eq:V2}
\end{equation}

Let us consider eq.\refer{eq:V2}
in a continuous and infinite plane wave (PW) basis
$\bracket{x}{k} =(\beta \Omega)^{-\frac{1}{2}} \exp(ikx)$, where
$\Omega$ is the volume. Then the matrix element
$\bra{k'}V\ket{k}$ equals
$\Omega^{-1}\,\delta_{q_0,0}V({\bf q})$,
i.e; the Fourier transform of $V({\bf r})$ with $q=k-k'$. Using the
dimensionless variables ${\bf x}={\bf r}/R$ and ${\bf t}={\bf q}R$
and keeping for $V$ only the first term in eq.\refer{eq:V} we obtain
for \refer{eq:V2} in the continuum limit \cite{RK}
\begin{equation}
\frac{1}{\beta}\Gamma_F^{(s.s)} = \frac{1}{4} N_C M^2 R \int
\frac{d^3t}{(2\pi)^3} \tr |L({\bf t})|^2
\phi\left(\left(\frac{{\bf t}}{R}\right)^2\right) + O(R^2)
\label{eq:V2n}
\end{equation}
with
\begin{equation}
\phi(q^2)=\int\frac{d^4 k}{(2\pi)^4} \left(\frac{1}
{(k^2+M^2)((k+q)^2+M^2)}-\frac{1}{(k^2+M^2)^2}\right)\\
\label{eq:phi}
%L({\bf t})&=&\int d^3x\,e^{i{\bf t}\cdot{\bf x}}\left(\fslash{\partial}_x
%\tilde {U}({\bf x})\right),
%\label{eq:L}
\end{equation}
and $L({\bf t})$ is the Fourier transform of $\left(\fslash{\partial}_x
\tilde {U}({\bf x})\right)$:
\begin{equation}
L({\bf t})=\int e^{i{\bf t}\cdot{\bf x}} \,\,\,\fslash{\partial}_x
\tilde {U}({\bf x}) d^3x.
\label{eq:four}
\end{equation}
Adding the kinetic term, we get
the leading contribution to the effective action in the small size expansion
as
\bea
\frac{1}{\beta}\Gamma^{(s.s)} = \frac{M^2 R}{16 g^2} \int
\frac{d^3t}{(2\pi)^3} \tr |L({\bf t})|^2
(1+4g^2N_C\phi\left(\left(\frac{{\bf t}}{R}\right)^2\right)) + O(R^2).
\label{eq:V2n.1}
\eea
Since
${\displaystyle \phi(q^2)\rightarrow -\frac{1}{16\pi^2} \ln\frac{q^2}
{M^2}}$ as $q^2\rightarrow \infty$ it follows that \refer{eq:V2n} behaves
as \\
$\alpha\, M^2 R\, \ln(MR)$ for $R\rightarrow 0$ with $\alpha>0$, which
overcomes the kinetic term (the $1$ in (\refer{eq:V2n.1})) for $R$ small
enough,
leading to a negative value for $\Gamma^{(s.s)}$.
This means that
for small sizes R the energy of this 'non-translational invariant vacuum'
can become lower than the energy of the translational invariant vacuum.
\par
To see that the Landau ghost in the meson propagators is responsible
for this vacuum instability, we use the fact that the effective action,
when expanded around the fields in the translational invariant vacuum
($\sigma=v,\,\,{\bold \pi}=0$), gives the inverse $\sigma$ and $\pi$
propagators
in the translational invariant vacuum as the coefficients of the terms
of second order in $s\equiv \sigma-v$ and ${\bold \pi}$. If we express
$V$ of eq. \refer{eq:V} in terms of $\sigma, {\bold \pi}$ using
${\displaystyle U=\frac{1}{v}(\sigma+i\gamma_5{\bold \pi}\cdot{\bold \tau})}$,
we see that these second order terms are completely
contained in \refer{eq:V2}.
Taking also the terms of second order in
$s, {\bold \pi}$ in the mesonic term in \refer{eq:gmes}, we obtain \cite{KA2}
\begin{equation}
\frac{1}{\beta}\Gamma^{(2)}=\frac{1}{2}\int \frac{d^3q}{(2\pi)^3}
(s({\bf q})s(-{\bf q})G_{\sigma}^{-1}({\bf q}^2)+
{\bold \pi}({\bf q})\cdot{\bold \pi}(-{\bf q})G_{\pi}^{-1}({\bf q}^2))
\label{eq:g2}
\end{equation}
with the inverse Euclidean Schwinger-Dyson propagators
\begin{eqnarray}
G_{\pi}^{-1}(q^2) &=& q^2\left(1+4 g^2 N_C \phi(q^2)\right),
\label{eq:proppi}\\
G_{\sigma}^{-1}(q^2) &=& q^2\left(1+4 g^2 N_C \phi(q^2)\right)+m_{\sigma}^2
+16 g^2 N_C M^2 \phi(q^2).
\label{eq:propsi}
\end{eqnarray}
In fig. \ref{fig:prop} we show the inverse pion
propagator \refer{eq:proppi} for $g=4$ by the dashed line.
The previously noted behaviour of $\phi(q^2)$
leads to the Landau ghost pole at Euclidean $q^2$, and as a consequence
\refer{eq:g2} can become more and more negative for large ${\bf q}^2$,
i.e; for small sized fields. In this limit of small sized fields the
leading contribution to \refer{eq:g2} comes from the terms $\propto q^2$ in
eqs. \refer{eq:proppi} and \refer{eq:propsi}, and if expressed in terms
of ${\bf t}={\bf q}R$ this gives again eq. \refer{eq:V2n}. [Note that
${\displaystyle tr |L({\bf t})|^2
=\frac{8}{v^2}{\bf t}^2\left({\tilde s}({\bf t})
{\tilde s}(-{\bf t})+{\tilde {\bold \pi}}({\bf t})\cdot{\tilde {\bold \pi}}
(-{\bf t})\right)}$ with ${\displaystyle {\tilde s}({\bf t})
=\frac{1}{R^3}s({\bf q})}$ the
Fourier transform of ${\tilde s}({\bf x})$ and similar for
${\tilde {\bold \pi}}({\bf t})$.]

\section{Ghost removal in a finite system}
We now address the problem of removing the Landau ghost in a finite
system. The procedure is described in ref. \cite{RED} and
has been extended to the path integral formalism in ref. \cite{TAN}.
To calculate the ground state energy we have to first obtain meson
propagators with the correct analytical properties. These propagators are
then implanted into the effective action which is in turn used to
calculate the ground state energy.

The prescription of ref. \cite{TAN} to eliminate the Landau ghost
from the effective
action is to replace the Schwinger-Dyson propagators $G_{\alpha}$
in the second order term \refer{eq:g2}
by the K\"{a}ll\'{e}n-Lehmann (KL) propagators
$\Delta_{\alpha}$. According to ref. \cite{RED}, these are
constructed from the KL representation
using a spectral function obtained from the one-loop meson self energy,
and are free of the Landau ghost. Thus, in this method the loop
approximation is is used only for the {\em imaginary} part, while the
real part is calculated from the dispersion relation. Since this method
avoids the Landau ghost, it clearly
represents an improvement of the straight forward loop approximation for
the whole propagator. The new effective action becomes
\begin{equation}
\Gamma_{KL}=\Gamma+\delta\Gamma, \,\,\,\,\,\,\,
\delta\Gamma\equiv \Gamma_{KL}^{(2)}-\Gamma^{(2)}.
\label{eq:gkl}
\end{equation}
In order to be consistent with chiral symmetry, however, this ghost
subtraction has to be done under the constraint of the Ward
identity \cite{GL}
$\Delta_{\sigma}(q^2)^{-1}-\Delta_{\pi}(q^2)^{-1}=ivT(-q;q,0)$
with $T(-q;q,0)$ the $\sigma\pi^2$ vertex where one of the external
pions has zero momentum. To preserve this identity the difference of the
$\sigma$ and $\pi$ inverse propagators has to remain invariant under
ghost subtraction \cite{KA2}, i.e; $\Delta_{\sigma}^{-1}-G_{\sigma}^{-1}=
\Delta_{\pi}^{-1}-G_{\pi}^{-1}$, and we obtain from \refer{eq:gkl} and
\refer{eq:dg}
\begin{equation}
\frac{1}{\beta}\delta\Gamma=\frac{1}{2}\int \frac{d^3q}{(2\pi)^3}
(s({\bf q})s(-{\bf q})+
{\bold \pi}({\bf q})\cdot{\bold \pi}(-{\bf q}))(\Delta_{\pi}^{-1}({\bf q}^2)-
G_{\pi}^{-1}({\bf q}^2)).
\label{eq:dg}
\end{equation}
It can be shown \cite{RED} that the KL propagator can be
obtained from the SD propagator
by subtracting the ghost pole:
$\Delta_{\pi}(q^2)=G_{\pi}(q^2)-Z_G/(q^2-m_G^2)$, where $Z_G<0$ is the residue
of the pole at Euclidean $q^2=m_G^2$. We show $\Delta_{\pi}^{-1}$ for
$m_{\pi}=0$ and $g=4$ by the solid line in fig.
\ref{fig:prop}.

The ghost free ground state energy is thus obtained by adding the piece
\refer{eq:dg} to $\frac{1}{\beta} \Gamma$ with $\Gamma$ given by eq.
\refer{eq:sum}. The exact evaluation of $\Gamma_F$, given by eq.
(\ref{eq:gferm}),
requires the diagonalization of  the Dirac hamiltonian
$h=-i{\bold \alpha}\cdot {\bold \nabla}+\beta_D MU$, and this is
done most conveniently in a basis $\ket{\lambda}$ which diagonalizes
the free Dirac
hamiltonian $h_0 = h(U=1) = -i{\bold \alpha}\cdot
{\bold \nabla}+\beta_D M$:
$h_0\ket{\lambda}=\epsilon^0_\lambda\ket{\lambda}$ with
$e^0_\lambda = \sqrt{{\bf k}^2_\lambda + M^2}$ in a box of size D with
discrete momenta $|{\bf k}_\lambda|<k_{max}$. (Here $\lambda$ labels
all necessary quantum numbers
except colour.) D and $k_{max}$ should be taken large
enough such that
all results are unchanged under further increase. We will call this
discrete and finite basis the Kahana-Ripka (KR) basis \cite{KR}.
On the other hand, the ghost subtraction is formulated most conveniently
in the continuous and infinite PW basis, for which the result
is given by eq. \refer{eq:dg}.
Therefore, we proceed in two steps: First, in order to demonstrate that
the KR basis gives the same results as the PW basis provided that
$k_{max}$ and $D$ are taken large enough, we evaluate $\Gamma_F$
in the small size approximation both in the KR and the PW basis and
compare the results. Then we evaluate $\Gamma_F$ exactly in the KR basis
and add the ghost subtraction term \refer{eq:dg}.

The expression for $\Gamma_F$ in the small size approximation using the
PW basis has already been given in eq. \refer{eq:V2n}, and the corresponding
result for the KR basis is obtained from eq. \refer{eq:V2} as
\begin{equation}
\frac{1}{\beta}\Gamma_F^{(s.s)}=\frac{1}{8} N_C \sum_{\lambda' \lambda}
|\bra{\lambda'}V\ket{\lambda}|^2
\left(\frac{1}{|e^0_{\lambda'}||e^0_\lambda|(|e^0_{\lambda'}|+
|e^0_{\lambda}|)}-\frac{1}{2}\frac{1}{|e^0_{\lambda'}|^3}\right).
\label{eq:V2nn}
\end{equation}
\par
Since for the exact calculation one has to diagonalize $h$, it is
convenient here to treat $\bra{\lambda'}V\ket{\lambda}$ as $h^2-h_0^2$ where
$h^2_{\alpha\beta} = h_{\alpha\gamma}h_{\gamma\beta}$ is the square
of a finite dimensional matrix.
In fig. \ref{fig:ss} we show the small-size approximation to the Fermion
loop energy, based on eq. \refer{eq:V2n} (solid line) and eq.
\refer{eq:V2nn} (dashed line) in units of $MN_C$
as a function of $MR$, assuming a Hedgehog profile with the vacuum value
for the chiral radius and winding number $n=1$:
\begin{equation}
U=exp(i{\hat{\bf r}}\cdot{\bold \tau}\Theta(r)\gamma_5),\,\,\,\,\,\,\,\,
\Theta(r)=\pi exp(-r/R).
\label{eq:prof}
\end{equation}
(In this case there are no additional $O(R^2)$
terms in \refer{eq:V2n} due to $\dag{U}U=1$. Also, $\Gamma_F/M$ or
$\Gamma_F^{(s.s)}/M$, viewed as a function of $MR$, is independent of
$M$ (or $g$). When the KR basis is used,
the effective action
becomes a function of $k_{max}$, and in order to reach convergence
($\Gamma(k_{max})\rightarrow \Gamma$) in the small-size approximation we
had to go up as high as $k_{max}\simeq 40 M$ for $MR\simeq 1$.
This is in contrast to models with
finite cut-off like the NJL model, where $k_{max}\simeq 10 M$ is
sufficient for solitons of normal size \cite{KA1}. The calculations in
this paper are
performed with values for $(k_{max}/M,DM)$ ranging from (70,5.7) for small
$R$ to (20,20) for large $R$, such that $k_{max}D\simeq 400$ is kept
constant.
Fig. \ref{fig:ss} demonstrates that the results using the PW and the KR
basis agree well if $k_{max}$ and also the box size $D$ are taken large
enough.
\par
Comparison with the full calculation, to be discussed below, shows that
the result of fig. \ref{fig:ss} is quite accurate up to $MR\simeq 0.7$.
The fact that the ground state energy relative to the translational
invariant vacuum is negative for these small R indicates the vacuum
instability due to the Landau ghost discussed above.
\par
The total ground state energy in the small size approximation is shown
in fig. \ref{fig:result}. It is obtained from eq. \refer{eq:V2n.1},
and for the ghost
free model by further adding the piece \refer{eq:dg}. We see that
the ground state energy in the small size approximation becomes a positive
quantity after removal of the Landau ghost.
\par
We now return to the full effective action,
eqs. \refer{eq:gferm},\refer{eq:gmes}. In the
KR basis $\Gamma_F$ can be written in the form \cite{RK}
\begin{equation}
\frac{1}{\beta}\Gamma_F=-\frac{N_C}{2} \sum_{\lambda}
\left(|\epsilon_{\lambda}|-|\epsilon^0_{\lambda}|\right)
+\frac{N_C}{4}\sum_{\lambda}\frac{\bra{\lambda}V\ket{\lambda}}
{|\epsilon^0_{\lambda}|}-
\frac{N_C}{16}\sum_{\lambda}\frac{\bra{\lambda}V^2\ket{\lambda}}
{|\epsilon^0_{\lambda}|^3}
\label{eq:gfermn}
\end{equation}
with $V=h^2-h_0^2$ as before, and $\epsilon_{\lambda}$ are the eigenvalues of
h.
Since we have already demonstrated for the small size expansion that
for $k_{max}$ and $D$ large enough the results with the KR basis agree
well with  those of the PW basis, it is feasible to use the expression
in the PW basis, eq. \refer{eq:dg}, for the ghost subtraction term
$\delta \Gamma$.\par
Fig. \ref{fig:result} shows the total ground state
energy $\frac{1}{\beta}\Gamma$
including the ghost (dashed line) and $\frac{1}{\beta}\Gamma_{KL}$
without the ghost (solid line)
in units of $MN_C$
as a function of $RM$ for the Hedgehog meson profile \refer{eq:prof}.
After ghost subtraction, the total ground
state energy relative to the
translational invariant vacuum is positive, which demonstrates that
amending the two-point functions such that they satisfy the KL
representation is sufficient to stabilize the vacuum against the
formation of arrays of strongly localized meson field configurations.

\section{Summary}
In effective quark-meson theories, which have been designed to model
QCD in the energy region of hadron masses, the Landau ghost appears in
the meson propagators at rather low energies of  $\simeq 1 GeV$. Its
presence is a real problem for these theories and leads to an instability
of the translational invariant vacuum. In this paper we have taken
the viewpoint that the Landau ghost should be brought under control
without the ad hoc introduction of further parameters.
Based on the work of Redmond and
Bogoliubov et al \cite{RED}, who have shown how one can
improve the loop expansion such
that the propagators have the correct analytical properties,
we have shown a practical way to eliminate the Landau ghost
from chiral solitons in the $\sigma$-model. We have demonstrated that the
ghost free effective action formalism of ref. \cite{TAN} can be
applied successfully to
finite systems. In particular, since the ghost subtraction term is
expressed most conveniently in a continuous and infinite plane wave basis,
we have investigated in detail the conditions under which it is feasible to add
this term to the Fermion loop term which is evaluated in a discrete
and finite basis. By considering vacuum configurations characterized by
various sizes of the meson profiles we have shown
that in our ghost free model the translational
invariant vacuum has the lowest energy and is therefore stable with
respect to decay into small sized configurations.
\par
The method given in this paper can form the base for further investigations
on self consistent soliton solutions in the $\sigma$-model. In particular,
as mentioned in the Introduction, it would be interesting to see whether
self consistent solutions can be obtained with a reasonable strength of
the $\phi^4$ term.
\par
The problem of the Landau ghost and the associated vacuum instability
occurs also in relativistic meson-nucleon theories for nuclear structure
\cite{SIOM}. For example, if a finite nucleus is described in the relativistic
Hartree approximation including the effect of the Dirac sea, one has to
apply the same 'overall ghost subtraction' as performed in this paper
(see eq. \refer{eq:gkl}). In more sophisticated approximations like the
'1/N' expansion \cite{EOS} or the 'modified loop expansion' \cite{WEHR}
the Landau ghost appears also in the subgraphs (even for infinite systems),
and the ghost subtraction must be applied to these subgraphs. Thus, although
we used the language of effective quark theories in this paper, the
method is applicable to relativistic nuclear structure physics as well.

\newpage

\section*{Figure caption}

\begin{enumerate}

\item{The inverse pion propagator with $m_{\pi}=0,\,\,g=4$
for Euclidean $q^2$. The dashed line shows
the Schwinger-Dyson propagator and the solid line the
K\"{a}ll\'{e}n-Lehmann propagator.}
\label{fig:prop}
\item{The Fermion loop energy $E_F=\frac{1}{\beta}\Gamma_F$ in
units of $MN_C$
as a function of $MR$ in the small size approximation. The solid line
shows the result with the PW basis, eq.\refer{eq:V2n}, and the dashed line
shows the result with the KR basis, eq. \refer{eq:V2nn} with values for
$k_{max}$ and $D$ described in the main text.}
\label{fig:ss}
\item{The total energy $E=\frac{1}{\beta}\Gamma$ in units of $M$
as a function of $MR$. The dotted
and dashed-dotted lines refer to the small size expansion before and after
the ghost subtraction, respectively, and the dashed and solid lines show the
full result before and after the ghost subtraction, respectively.}
\label{fig:result}

\end{enumerate}


\newpage

\begin{thebibliography}{99}
\bibitem{EFF} E. Witten, Nucl. Phys. {\bf B 233} (1983) 422; 433 \\
              H.B. Nielsen and A. Pathkos, Nucl. Phys. {\bf B 195}
              (1982) 137 \\
              M. Betz and R. Goldflam, Phys. Rev. {\bf D 28} (1983) 2848.
\bibitem{GL} M. Gell-Mann and M. L\'{e}vy, Nuovo Cimento {\bf 16}
(1960) 705\\
    B.W. Lee, Chiral Dynamics, Gordon and Breach, New York, 1972.
\bibitem{ACT} T. Eguchi, Phys. Rev. {\bf D 14} (1976) 2755 \\
              I.J. Aitchison and C.M. Fraser, Phys. Lett. {\bf B 146}(1984)
              63; Phys. Rev. {\bf D 31} (1985) 2605.
\bibitem{BR} W. Broniowski and M. Kutschera, Phys. Lett. {\bf B 242} (1990)
             133.
\bibitem{REN} J. Mignaco and E. Remiddi, Nuov. Cim. {\bf 1 A} (1971) 376.
\bibitem{LAN} L. D. Landau, in Niels Bohr and the development of physics,
ed. W. Pauli (Pergamon, London, 1955).
\bibitem{RK} G. Ripka and S.Kahana, Phys. Rev. {\bf D 36} (1987) 1233.
\bibitem{VAC} R. Perry, Phys. Lett. {\bf B 199} (1987) 489 \\
V. Soni, Phys. Lett. {\bf B 183} (1987) 91 \\
T.D. Cohen, M.K. Banerjee and V-Y. Ren, Phys. Rev. {\bf C 36} (1987) 1653 \\
R.J. Furnstahl and C.J. Horowitz, Nucl. Phys. {\bf A 485} (1988) 632 \\
T. Kohmura, Y. Miyama, T. Nagai, S. Ohnaka and J.Da Provid\^{e}ncia,
 Phys. Lett. {\bf B 226} (1989) 207 \\
T. Mannel, T. Ohl and P. Manakos, Z. Phys. {\bf A 335} (1990) 341.
\bibitem{RPA} S. Ichii, W. Bentz, A. Arima and T. Suzuki, Phys. Lett.
              {\bf 192} (1987) 11.
\bibitem{RED} P.J. Redmond, Phys. Rev. {\bf 112} (1958) 1404;\\
    N.N. Bogoliubov, A.A. Logunov and D.V. Shirkov, Sov. Phys. JETP
    (Engl. Transl.) {\bf 37} (1960) 574.
\bibitem{DYS} F.J. Dyson, Phys. Rev. {\bf 85} (1952) 631.
\bibitem{TAN} K. Tanaka, W. Bentz, A. Arima, F. Beck, Nucl. Phys. {\bf A 528}
(1991) 676\\
   K. Tanaka and W. Bentz, Nucl. Phys. {\bf A 540} (1992) 383.
\bibitem{KA2} M. Kato, W. Bentz and K. Tanaka, Phys. Rev. {\bf C 45}
(1992) 2445.
\bibitem{GAUGE} S.H. Kahana and G. Ripka, Phys. Lett. {\bf 278} (1992)
   11.
\bibitem {NJL} Y. Nambu and G. Jona-Lasinio, Phys. Rev. {\bf 122} (1961) 345,
               {\bf 124} (1961) 246 \\
      U. Vogl and W. Weise, Prog. Part. Nucl. Phys. {\bf 27} (1991) 195.
\bibitem{SOLI} P. Sieber, Th. Meissner, F. Gr\"{u}mmer and K. Goeke, Nucl.
       Phys. {\bf A547} (1992) 459 \\
       T. Watabe and H. Toki, Progr. Theor. Phys. {\bf 87} (1992) 651 \\
       J. Schlienz, H. Weigel, H. Reinhardt and R. Alkhofer,
       Phys. Lett. {\bf B 315} (1993) 6.
\bibitem{KA1} M. Kato, W. Bentz, K. Yazaki and K. Tanaka, Nucl. Phys.
{\bf A 551} (1993) 541.
\bibitem{ISH} N. Ishii, W. Bentz and K. Yazaki, Phys. Lett. {\bf B 301} (1993)
              165; {\bf B 318} (1993) 26.
\bibitem{PHI} Th. Meissner, G. Ripka, R. W\"{u}nsch, P. Sieber,
F. Gr\"{u}mmer and K. Goeke, Phys. Lett. {\bf B 299} (1993) 183\\
C. Weiss, R. Alkhofer and H. Weigel, Mod. Phys. Lett. {\bf A 8} (1993) 79.
\bibitem{KR} S. Kahana and G. Ripka, Nucl. Phys. {\bf A 429} (1984) 462.
\bibitem{SIOM} B.D. Serot and J.D. Walecka, Adv. in Nucl. Phys. {\bf 16},
               ed. J.W. Negele and E. Vogt, Plenum, 1986,1 \\
               B.D. Serot, Rep. Prog. Phys. {\bf 55} (1992) 1855.
\bibitem{EOS}  K. Tanaka, W. Bentz and A. Arima, Nucl. Phys. {\bf A 555}
               (1993) 151.
\bibitem{WEHR} K. Wehrberger, R. Wittman and B.D. Serot. Phys. Rev.
               {\bf C 42} (1990) 2680.


\end{thebibliography}
\end{document}